\documentclass[conference]{IEEEtran}
\usepackage{amsmath}
\usepackage{amsfonts}
\usepackage{cite}
\usepackage{array}
\usepackage{graphicx}

\newcommand{\latin}[1]{\textit{#1}}
\newcommand{\term}[1]{\textit{#1}}

\newcommand{\set}[1]{\ensuremath{\mathcal{#1}}}

\newcommand{\vect}[1]{\ensuremath{\mathbf{#1}}}

\newcommand{\goesto}{\ensuremath{\to}}

\newcommand{\dvavg}{\ensuremath{\overline{d}_v}}

\renewcommand{\ne}{\ensuremath{n_{\mathfrak{e}}}}
\newcommand{\nr}{\ensuremath{n_{\mathfrak{r}}}}

\newcommand{\etal}{\latin{et al}.\ }
\newcommand{\etalnsp}{\latin{et al}.}

\newcommand{\secref}[1]{Section~\ref{#1}}
\newcommand{\appref}[1]{Appendix~\ref{#1}}
\newcommand{\figref}[1]{Fig.~\ref{#1}}

\renewcommand{\P}{\ensuremath{\text{P}}}

\newcommand{\bmat}[2][c]{\ensuremath{\left[\begin{array}[#1]{#2}}}
\newcommand{\emat}{\ensuremath{\end{array}\right]}}

\newcommand{\diff}{\ensuremath{\mathrm{d}}}
\newcommand{\intd}{\ensuremath{\mathrm{d}}}
\newcommand{\doverd}[2][]{\ensuremath{\frac{\diff{}#1}{\diff{}#2}}}
\newcommand{\ddt}{\ensuremath{\frac{\diff}{\diff t}}}

\newcommand{\partialwrt}[2][]{\ensuremath{\frac{\partial #1}{\partial #2}}}

\newcommand{\va}{\ensuremath{\vect{a}}}
\newcommand{\vb}{\ensuremath{\vect{b}}}

\newcommand{\vd}{\ensuremath{\vect{d}}}
\newcommand{\ve}{\ensuremath{\vect{e}}}

\newcommand{\vr}{\ensuremath{\vect{r}}}

\newcommand{\vx}{\ensuremath{\vect{x}}}

\newcommand{\vone}{\ensuremath{\boldsymbol{1}}}

\newcommand{\vepsilon}{\ensuremath{\boldsymbol{\epsilon}}}

\newcommand{\vlambda}{\ensuremath{\boldsymbol{\lambda}}}

\newcommand{\vrho}{\ensuremath{\boldsymbol{\rho}}}

\makeatletter
\let\preamsthmproof\proof
\let\preamsthmendproof\endproof
\let\proof\@undefined
\let\endproof\@undefined
\makeatother
\usepackage{amsthm}
\let\proof\preamsthmproof
\let\endproof\preamsthmendproof

\newtheorem{lemma}{Lemma}

\ifdefined\IEEEQED
\else
  \newcommand{\IEEEauthorblockA}{\authorblockA}
  \newcommand{\IEEEauthorblockN}{\authorblockN}

  \newcommand{\IEEEQED}{\QED}
  
  \newcommand{\IEEEQEDopen}{\QEDopen}
  
\fi
\renewcommand{\IEEEQED}{\IEEEQEDopen}

\newcommand{\dvidxavg}[1]{\ensuremath{\overline{d}_{v,#1}}}
\newcommand{\dvavgInv}{\ensuremath{\frac{1}{\,\dvavg\,}}}

\newcommand{\Dt}{\ensuremath{\Delta t}}
\newcommand{\LDPCens}{\ensuremath{\mathrm{LDPC}}}
\newcommand{\weight}[1]{\ensuremath{\lVert #1 \rVert}}

\IEEEoverridecommandlockouts  
\title{Analysis of peeling decoder for MET ensembles}%
\author{%
  \IEEEauthorblockN{Ryan W. Hinton}%
  \IEEEauthorblockA{%
    University of Virginia and L-3 Communications CSW \\%
    rwh4s@virginia.edu}%
  \and%
  \IEEEauthorblockN{Stephen G. Wilson}%
  \IEEEauthorblockA{%
    University of Virginia \\%
    sgw@virginia.edu}%
  \thanks{This work was supported in part by the National Science Foundation
    under grant number ECCS 0636598.}%
}%

\begin{document}
\maketitle

\begin{abstract}
  The peeling decoder introduced by Luby, \etal allows analysis of LDPC
  decoding for the binary erasure channel (BEC).  For irregular ensembles, they
  analyze the decoder state as a Markov process and present a solution to the
  differential equations describing the process mean.  Multi-edge type (MET)
  ensembles allow greater precision through specifying graph connectivity.  We
  generalize the the peeling decoder for MET ensembles and derive analogous
  differential equations.  We offer a new change of variables and solution to
  the node fraction evolutions in the general (MET) case.  This result is
  preparatory to investigating finite-length ensemble behavior.
\end{abstract}

\section{Introduction}
Low-density parity-check codes (LDPC) offer excellent channel coding
performance using a simple decoding algorithm\cite{Chung2001}.  Luby,
\etalnsp\cite{Luby1997} introduced the \term{peeling decoder} as a tool for
analyzing irregular ensembles over the binary erasure channel (BEC).  With this
tool they obtain a form of \term{density evolution} and conditions to determine
\term{threshold}, the asymptotic limit of the ensemble decoding performance.

Multi-edge type (MET) LDPC ensembles\cite{Richardson2002} generalize irregular
ensembles by allowing control over graph connectivity.  In particular, the MET
framework allows for degree-1 variable nodes, punctured variable nodes, and
control over the graph structure that offer superior performance relative to
irregular ensembles of comparable complexity (block length, average and maximum
degree).  In addition, MET ensembles are useful analytically as they include
ensembles with interesting structure such as repeat-accumulate, irregular
repeat-accumulate, and protograph codes as special cases---as well as regular
and irregular ensembles.

In this paper we generalize the peeling decoder analysis to multi-edge type
ensembles.  Specifically, we present the modified difference equations and
corresponding differential equations, introduce a new change of variables, and
solve the system.  The original MET analysis\cite{Richardson2002} treats belief
propagation for the larger class of symmetric binary, memoryless, symmetric
channels.  However, the Markov chain setting is more easily applied to
finite-length performance scaling\cite{Amraoui2009}.  We plan to build on this
result to develop finite-length scaling laws for MET ensembles over the BEC and
eventually for general channels as in~\cite{Ezri2007}.

The remainder of the paper is organized as follows.  \secref{sec:background}
provides background on and notation for the peeling decoder and multi-edge type
ensembles.  \secref{sec:met-pd} describes the system of difference and
ordinary differential equations (ODEs) for the peeling decoder applied to MET
ensembles.  \secref{sec:met-pd-soln} presents a new change of variables and the
corresponding ODE system solution.  \secref{sec:met-xpath} discusses this
solution and concludes the paper.

\section{Background}
\label{sec:background}
\begin{figure*}[t]
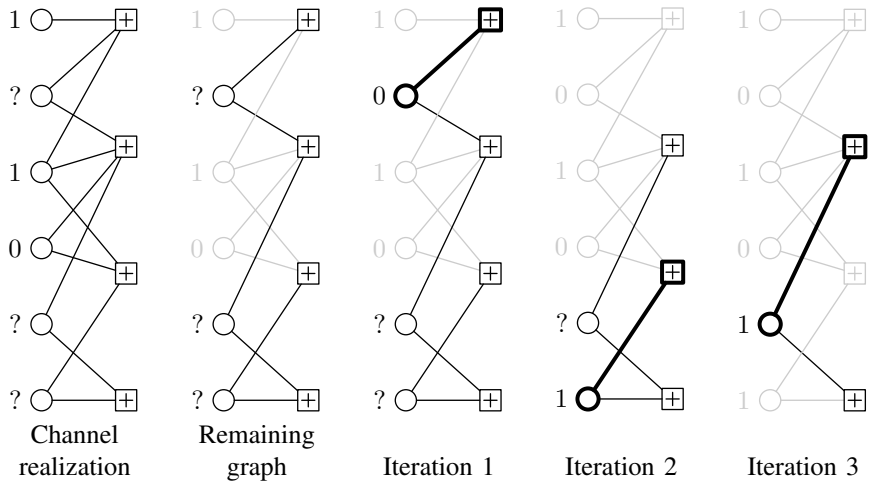

  \centering
  \begin{tabular}{>{\centering{}}b{2.0cm}>{\centering{}}b{2.0cm}>{\centering{}}b{2.0cm}%
                  >{\centering{}}b{2.0cm}>{\centering{}}b{2.0cm}c}
    \includegraphics{proposal-21.mps}  &
    \includegraphics{proposal-22.mps}  &
    \includegraphics{proposal-23.mps}  &
    \includegraphics{proposal-24.mps}  &
    \includegraphics{proposal-25.mps}  & \\
    Channel \linebreak{}realization & Remaining graph & Iteration 1 & 
      Iteration 2 & Iteration 3 & 
  \end{tabular}
  \caption{Peeling decoder operation.}
  \label{fig:peeling-decoder}
\end{figure*}

\subsection{Multi-edge type ensembles}
\label{sec:bg-met}
An MET ensemble\cite{Richardson2002} $\LDPCens(N, \nu, \mu)$ is
specified by the transmitted block length $N$ and a pair of multivariate
generating functions, $\nu(\vr,\vx)$ and $\mu(\vx)$
\begin{align*}
  \nu(\vr,\vx) &= \sum_{\vb,\vd} \nu_{\vb,\vd} \vr^{\vb} \vx^{\vd}  &
  \mu(\vx) &= \sum_{\vd} \mu_{\vd} \vx^{\vd}  \\
  \vd &= (d_1, d_2, \dots, d_{\ne})  &
  \vb &= (b_0, b_1, \dots, b_{\nr})  \\
  \vx &= (x_1, x_2, \dots, x_{\ne})  &
  \vr &= (r_0, r_1, \dots, r_{\nr})  
\end{align*}
where $\vx^{\vd} = \prod_{i=1}^{\ne} x_i^{d_i}$ and $\vr^{\vb} =
\prod_{i=0}^{\nr} r_i^{b_i}$.  Note that the edge type index (i.e. the
subscript $i$ on $x_i$) goes from $1$ to $\ne$ while the channel type index
(i.e. subscripts of $r_i$) goes from $0$ to $\nr$ in order to include the
punctured channel\footnote{Punctured data can be considered as transmitted through
  an erasure channel with erasure probability 1.}, $r_0$.  Each term
($\nu_{\vb,\vd}\vr^{\vb}\vx^{\vd}$) in $\nu$ corresponds to variable nodes of a
single type: the coefficient $\nu_{\vb,\vd}$ is the fraction of nodes of type
$(\vb,\vd)$, the exponent $(b_i)$ on each channel type variable $(r_i)$
indicates the number of connections to that channel\footnote{The channel type
  exponents $\vb$ typically have a single non-zero entry of $1$ indicating that
  each bit receives exactly one channel realization.}, and the exponent $(d_j)$
on each edge type variable $(x_j)$ indicates the number of sockets of that
type.  The corresponding definitions hold for the coefficients $\mu_{\vd}$ and
edge type exponents $\vd$ of the terms of $\mu$.  In particular, note that
variable nodes have type or degree $(\vb,\vd)$ to distinguish their channel and
edge connectivity while check nodes are distinguished solely by their edge
connectivity, $\vd$.
As with irregular ensembles, the ensemble consists of the codes corresponding
to every (compatible) socket permutation.  

The remaining ensemble characteristics can be derived from these polynomials.
The number of edges of type $i$ connected to a particular variable node or
check node type is given by $Nd_i\nu_{\vb,\vd}$ and $Nd_i\mu_{\vd}$,
respectively.  Define
\begin{align*}
  \nu_{x_i}(\vr,\vx) &= \partialwrt{x_i} \nu(\vr,\vx)  &
  \mu_{x_i}(\vr,\vx) &= \partialwrt{x_i} \mu(\vr,\vx)  
\end{align*}
and the total number of edges of type $i$ as
\begin{align*}
  E_i = N\nu_{x_i}(\vone,\vone) = N\mu_{x_i}(\vone)
\end{align*}
and $E=\sum_i E_i$ the total number of edges.  The edge-perspective degree
profiles are
\begin{align*}
  \vlambda(\vr,\vx) 
    &= \left(\frac{\nu_{x_1}(\vr,\vx)}{\nu_{x_1}(\vone,\vone)}, \dots, 
      \frac{\nu_{x_{\ne}}(\vr,\vx)}{\nu_{x_{\ne}}(\vone,\vone)} \right)  \\
  \vrho(\vx) 
    &= \left(\frac{\mu_{x_1}(\vx)}{\mu_{x_1}(\vone)}, \dots, 
      \frac{\mu_{x_{\ne}}(\vx)}{\mu_{x_{\ne}}(\vone)} \right)
\end{align*}
where the denominators can also expressed as $E_i/N$.  



\subsection{Peeling decoder}
\label{sec:bg-pd}
The peeling decoder of~\cite{Luby1997} attempts to recover an LDPC code word
transmitted over the erasure channel.  Its operation is depicted
in \figref{fig:peeling-decoder}.  Given the channel realization (i.e. which
bits were erased by the channel), each check node sums the known values from
its attached variable nodes.  Then the known variable nodes and their edges are
deleted from the decoding (Tanner) graph since they convey no further
information.  
The check node constraint allows us to deduce the value of a variable node
connected to a check node with degree 1 in the remaining graph.  Decoding
proceeds iteratively on this principle.  At each iteration $T=1,2,\dots$, a
degree-1 check node is selected at random.  The attached variable node value is
computed, then this node and the selected check node are deleted along with
their edges.  Decoding stops when no degree-1 check nodes remain.  If the
remaining graph is empty, decoding was successful.  Otherwise, the remaining
variable nodes form a \term{stopping set}\cite{Richardson2001a,Di2002}.  The
critical insight is that the distribution of socket permutations for the
remaining graph at each iteration \emph{remains} uniform, so the evolution of
the degree sequence can be described as a Markov chain of degree distributions.

\section{Peeling decoder for MET ensembles}
\label{sec:met-pd}
Consider the operation of the peeling decoder on a code from an MET ensemble.
This development closely follows~\cite{Luby1997}.  However, we examine the
evolution of node fractions instead of edge fractions: allowing multiple edge
types results in a proliferation of redundant edge-perspective equations.

Analyzing the peeling decoder for MET ensembles requires extending and adding
notation.  The node fractions are normalized by $N$ while the edge fractions
are normalized by $E$, so the average variable node degree $\dvavg=E/N$ and
per-edge type average degrees $\dvidxavg{i}=E_i/N$ frequently appear as
conversion factors.  Let $\ve_i$ denote the unit vector with a $1$ in position
$i$ and zeros elsewhere.  By considering only the erasure channel, the vector
of erasure probabilities $\vepsilon=(1,\epsilon_1,\epsilon_2,\dots)$ can be
substituted for the general distributions $\vr$.

Consider the decoder state at time $t=T/E$.  Define the time functions
$\nu_{\vb,\vd}(t)$ to indicate the remaining fraction (relative to $N$) of
variable nodes of each type $(\vb,\vd)$ at time $t$ and similarly for check
nodes.  These time-varying parameters are not to be confused with
$\nu_{\vb,\vd}$ and $\mu_{\vd}$, the fraction of nodes of each type in the
Tanner graph before decoding.  Denote $\ell_{i,\vb,\vd}(t)=\dvavg
d_i\nu_{\vb,\vd}(t)$ the remaining fraction of edges (relative to $E$) of type
$i$ connected to variable nodes of type $(\vb,\vd)$ at time $t$ and
$r_{i,\vd}(t)=\dvavg d_i \mu_{\vd}(t)$ the analogous check node edge fraction.
Summing either set gives
\begin{align*}
    e_i(t) &= \sum_{\vb,\vd} \ell_{i,\vb,\vd}(t) 
     = \sum_{\vd} r_{i,\vd}(t),
\end{align*}
the remaining fraction of edges of type $i$ at time $t$.  Denote
$N_{\vb,\vd}(t)=N\nu_{\vb,\vd}(t)$ the expected number of variable nodes of
type $(\vb,\vd)$ at time $t$ and $M_{\vd}(t)=N\mu_{\vd}(t)$ the analogous check
node mean.  Let the time step $\Dt=1/E$.  Finally, denote the indicator
function
\begin{align*}
  I\{x\} = \begin{cases} 1 & \text{if $x$ is true} \\
    0 & \text{otherwise}.
  \end{cases}
\end{align*}

At each successful iteration, the peeling decoder deletes a check node of
degree one, attached variable node, and their edges.  For a code from a
multi-edge type ensemble, there may be degree-one check nodes of several types
from which to choose.  Define the time-indexed sequence of random variables
$\Gamma(t)$ whose pmfs $\gamma_i(t)=\P(\Gamma(t)=i)$ indicate the probability
of choosing a degree-$\ve_i$ check node at each iteration.  Interestingly,
$\Gamma(t)$ presents a free variable for the decoder of $\ne-1$ dimensions at
each iteration.  Choosing $\Gamma(t)$ is discussed in~\secref{sec:met-xpath}.
Assuming the decoder chooses a degree-1 check node (CN) of type $i$, the
probability that the attached variable node (VN) has type $(\vb,\vd)$ is
\begin{align*}
  \frac{\ell_{i,\vb,\vd}(t)}{e_i(t)} = \frac{\dvavg d_i \nu_{\vb,\vd}(t)}{e_i(t)}.
\end{align*}
This value times $\gamma_i(t)$ gives the (joint) probability of choosing an edge
of type $i$ attached to a variable node of type $(\vb,\vd)$.  Hence the marginal
\begin{align*}
  \sum_{i=1}^{\ne} \gamma_i(t) \frac{\ell_{i,\vb,\vd}(t)}{e_i(t)} 
    &= \dvavgInv\nu_{\vb,\vd}(t) \sum_i \frac{d_i\gamma_i(t)}{e_i(t)}
\end{align*}
gives the total probability of removing a variable node of type $(\vb,\vd)$ at
time $t$.  The difference in the expected number of nodes of each type is
\begin{align}
  \label{eqn:diffN}
  N_{\vb,\vd}(t+\Dt) - N_{\vb,\vd}(t) 
    &= -\dvavgInv \nu_{\vb,\vd}(t) \sum_i \frac{d_i\gamma_i(t)}{e_i(t)}.
\end{align}

Having selected a degree-$\ve_i$ check node attached to a variable node of type
$(\vb,\vd)$, the decoder removes these nodes and their edges from the graph.
Removing the variable node's edges changes the degree or type of the connected
check nodes.  One edge connects to the original degree-1 check node, but we
need to account for the other $\vd-\ve_i$ edges.  The expected number of
``other'' edges of type $j$ deleted is%
\footnote{In the first expression, the number of other edges, $d_j-I\{i=j\}$,
  may go negative.  However, this is not a concern since this only happens when
  $d_j=0$ so $\ell_{i,\vb,\vd}(t)$ is zero.}
\begin{align*}
  &\hspace*{-0.1em}
      \sum_{i,\vb,\vd} (d_j - I\{i=j\}) 
      \gamma_i(t) \frac{\ell_{i,\vb,\vd}(t)}{e_i(t)}  \\
    &= \left[\dvavgInv \sum_{\vb,\vd} d_j \nu_{\vb,\vd}(t) 
        \sum_i \frac{d_i\gamma_i(t)}{e_i(t)}\right]
      - \left[\frac{\gamma_j(t)}{e_j(t)} 
        \cdot \sum_{\vb,\vd} \ell_{j,\vb,\vd}(t)\right]  \\
    &= \tilde{a}_j(t) -\gamma_j(t)
\end{align*}
where 
\begin{align}
  \label{eqn:expected-edges-deleted}
  \tilde{a}_j(t) &= \dvavgInv \sum_{\vb,\vd} d_j \nu_{\vb,\vd}(t) 
      \sum_i \frac{d_i\gamma_i(t)}{e_i(t)}  
\end{align}
is the expected (total) number of edges of type $j$ deleted at time $t$
.  
Assume each ``other'' edge is connected to an unique check node since the
computation graph is tree-like in the large block length limit.
The probability that a deleted variable node edge of type $j$ is connected
to a check node of type $\vd$ is $r_{j,\vd}(t) /
e_j(t)=d_j\mu_{\vd}(t)/\bigl[\dvavg e_j(t)\bigr]$.  Deleting an edge of type
$j$ from a check node of type $\vd+\ve_j$ removes one node of this type and
adds a node of type $\vd$.  So the expected number of added check nodes of type
$\vd$ is
\begin{equation*}
  \sum_j \bigl[\tilde{a}_j(t)-\gamma_j(t)\bigr] 
      \frac{(d_j+1)\mu_{\vd+\ve_j}(t)}{\dvavg e_j(t)}
\end{equation*}
and the expected number of removed check nodes of type $\vd$ is 
\begin{equation*}
   \gamma_i(t)I\{\vd\text{ is degree-1}\} 
      + \sum_j \bigl[\tilde{a}_j(t)-\gamma_j(t)\bigr] 
        \frac{d_j\mu_{\vd}(t)}{\dvavg e_j(t)}
.
\end{equation*}
The expected change in the number of check nodes of type $\vd$ is simply
their difference,
\begin{align}
  \label{eqn:diffM}
  &\hspace*{-2em}
  M_{\vd}(t+\Dt) - M_{\vd}(t) \notag\\
    &= \sum_j \bigl[(d_j+1)\mu_{\vd+\ve_j}(t) - d_j\mu_{\vd}(t)\bigr]
      \frac{\tilde{a}_j(t)-\gamma_j(t)}{\dvavg e_j(t)}  \notag\\
    &\qquad  - \sum_i \gamma_i(t) I\{\vd = \ve_i\}.
\end{align}

The system of difference equations given by \eqref{eqn:diffN} and
\eqref{eqn:diffM} describe the evolution of the expected number of variable and
check nodes of each type.  These quantities also provide the transition
probabilities for the Markov chain whose state is these node counts.  Since the
time axis $t$ is scaled inversely with $E$, in the large block length limit
$\Dt\goesto 0$ and the discrete-time Markov chain approaches a continuous-time
Markov process (see~\cite{Amraoui2009} for a formal description).  The
transition rates for this process are the limits of the corresponding
difference equations.  Since $N_{\vb,\vd}(t) = N\nu_{\vb,\vd}(t) =
E\nu_{\vb,\vd}(t)/\dvavg = \nu_{\vb,\vd}/(\dvavg \Dt)$, the limiting variable
node rates are
\begin{align}
\label{eqn:dnu}
  \ddt \nu_{\vb,\vd}(t)
    &= -\nu_{\vb,\vd}(t) \sum_i \frac{d_i\gamma_i(t)}{e_i(t)}.
\end{align} 
Similarly, the check node fractions follow
\begin{align}
  \label{eqn:dmu1}
  \ddt \mu_{\vd}(t) 
    &= \sum_j \bigl[(d_j+1)\mu_{\vd+\ve_j}(t) - d_j\mu_{\vd}(t)\bigr]
      \frac{\tilde{a}_j(t)-\gamma_j(t)}{e_j(t)} \notag\\
    &\qquad  - \dvavg\sum_i \gamma_i(t) I\{\vd = \ve_i\}.
\end{align}

\subsection{Initial conditions}
A complete solution requires the process initial conditions as well.  The known
nodes are deleted after the channel realization, so variable nodes attached to
channel type $k$ are retained with probability $\epsilon_k$.  Accordingly, the
initial means are
\begin{align}
  \label{eqn:nu-initial}
  \nu_{\vb,\vd}(0) &= \vepsilon^{\vb}\nu_{\vb,\vd}.
\end{align}
In other words, each variable node is \emph{deleted} from the remaining graph
according to the probability that it is \emph{not erased} over its attached
channel.

The check node fractions are more interesting.  The probability that an edge of
type $i$ is retained before the first iteration is
\begin{align*}
  \frac{\sum_{\vb,\vd} d_i\vepsilon^{\vb}\nu_{\vb,\vd}}%
                             {\sum_{\vb,\vd} d_i\nu_{\vb,\vd}}
    = \frac{\sum_{\vb,\vd} d_i\nu_{\vb,\vd}(0)}{E_i/N} 
    = \lambda_i(\vepsilon,\vone),
\end{align*}
the weighted average of the probabilities that the variable nodes attached to
this edge type will be retained.  A check node type may be produced by deleting
edges from any initial check node type with larger degree, so
\begin{align}
  \label{eqn:mu-initial}
  \mu_{\vd}(0) 
    &= \sum_{\tilde{\vd} \geq \vd} \mu_{\tilde{\vd}} \binom{\tilde{\vd}}{\vd}
      \vlambda(\vepsilon,\vone)^{\vd}
      \bigl[\vone - \vlambda(\vepsilon,\vone)\bigr]^{\tilde{\vd} - \vd}
\end{align}
where $\tilde{\vd} \geq \vd$ when $\tilde{d}_i \geq d_i$ for $i=1,\dots,\ne$
and the vector binomial denotes the product of binomial coefficients,
\begin{align*}
  \binom{\tilde{\vd}}{\vd} = \prod_i \binom{\tilde{d_i}}{d_i}.
\end{align*}
Note that the only non-zero terms in \eqref{eqn:mu-initial} are those with
positive fractions in the original Tanner graph, $\{\tilde{\vd} :
\mu_{\tilde{\vd}} > 0\}$.

\section{Solution for MET peeling decoder evolution}
\label{sec:met-pd-soln}
The ODE system solution for multi-edge types is similar to that for the single
edge case.  One key novelty is the change of variables from $t$ to
$\vx=(x_1,x_2,\dots,x_{\ne})$ with $x_i$ defined implicitly by
\begin{align}
  \label{eqn:gamma_e_x}
  \frac{\gamma_i(t)}{e_i(t)} &= -\frac{1}{x_i} \cdot \doverd[x_i]{t}  
\end{align}
This substitution conflates time with choice of $\Gamma(t)$.  For clarity, we
will refer to $\vx$ and its elements without making either dependence
explicit.  Using this substitution, the variable node differential equation is
\begin{align*}
  \ddt \nu_{\vb,\vd}(\vepsilon,\vx) 
    &= \nu_{\vb,\vd}(\vepsilon,\vx) \sum_i \frac{d_i}{x_i} \cdot \doverd[x_i]{t}
\end{align*}
with solution 
\begin{align}
  \label{eqn:nu-soln}
  \nu_{\vb,\vd}(\vepsilon,\vx) &= \nu_{\vb,\vd} \vepsilon^{\vb} \vx^{\vd}
\end{align}
including the initial condition \eqref{eqn:nu-initial} at $\vx=\vone$.

The more involved check node derivation is presented in
\appref{app:met-cn-soln}, but we include the result for degree-1 check nodes
here.
\begin{lemma}
  \label{lem:mu1-soln}
  The solution to the system~\eqref{eqn:dmu1},~\eqref{eqn:mu-initial} for
  degree-1 check nodes is given by
  \begin{align}
    \label{eqn:mu1-soln}
    \mu_{\ve_i}(\vepsilon,\vx) &=
      \nu_{x_i}(\vepsilon,\vx)\bigl[x_i - 1 + 
        \rho_i\bigl(\vone - \vlambda(\vepsilon,\vx)\bigr)\bigr].
  \end{align}
\end{lemma}
Note that this solution is equivalent to the the result from~\cite{Luby1997}
for the single-edge case.

\section{Discussion of solution and $\vx$-space path}
\label{sec:met-xpath}
Equation \eqref{eqn:mu1-soln} provides the fraction of degree-1 check nodes of
each type for any choice of \term{schedule}, $\Gamma(\cdot)$.  The node
fraction solutions are defined over the entire space $\set{X}=[0,1]^{\ne}$, but
the decoder behavior is modeled by these equations evaluated over the path
$\vx(t)$ induced by the schedule.
In particular, the natural choice is for the peeling decoder to select a
degree-1 check node without respect to edge type, i.e. according to the
proportion of remaining degree-1 nodes of that type:
\begin{align*}
  \gamma_i(\vepsilon,\vx) 
    = \frac{\mu_{\ve_i}(\vepsilon,\vx)}{\sum_j \mu_{\ve_j}(\vepsilon,\vx)}.
\end{align*}
This choice yields another system of differential equations for $\vx(t)$ which
we have not yet solved.  

For an ensemble with a single channel type (besides perhaps the punctured
channel), define the \term{schedule threshold} for a particular schedule as
\begin{align*}
  \epsilon_\Gamma^* &= \sup \{\epsilon : \mu_{\ve_i}\bigl(\vx(t)\bigr) > 0 
      \;\forall\; i=1,\dots,\ne \text{ and } t\in[0,t_f)\}
\end{align*}
where $\vx(t)$ denotes the path induced by the schedule $\Gamma$ and
$t_f=\nu(\vepsilon,\vone)/\dvavg$ is the expected decoder completion time.
Also define the \term{ensemble threshold} as
\begin{align*}
  \epsilon^* &= \max_{\Gamma} \{ \epsilon_\Gamma^* \}.
\end{align*}
In fact, the choice of schedule is not critical.  A \term{reasonable} schedule
always chooses an edge type with a positive fraction of degree-1 check
nodes when one is available.  For example, choosing $\gamma_i(t)=I\{i=0\}1$ is
an unreasonable schedule for a two-edge type code since the decoder will fail
when the supply of degree-$\ve_1$ check nodes is exhausted even though
degree-$\ve_2$ check nodes may be available.
\begin{lemma}
  Every reasonable schedule $\Gamma$ has $\epsilon_\Gamma^*=\epsilon^*$.  In
  other words, every reasonable schedule achieves the ensemble threshold.
\end{lemma}
\begin{IEEEproof}
  Failure events on the BEC are given by stopping sets\cite{Di2002}.  A
  reasonable schedule will fail only at a stopping set.  Since no schedule can
  decode past the unique maximal stopping set, every reasonable schedule must
  yield the same (optimal) error pattern.
\end{IEEEproof}

By standard arguments~\cite{Luby1997}, the behavior of codes from an MET
ensemble concentrate in probability around the means~\eqref{eqn:nu-soln},
\eqref{eqn:mu1-soln} and~\eqref{eqn:mu-soln} as the block lengh increases.  For
$\epsilon < \epsilon^*$, all of the degree-1 check node fractions remain
positive with high probability until decoding succeeds ($t=t_f$).  Likewise for
$\epsilon > \epsilon^*$, the decoder fails with high probability.

\begin{figure}[t]
  \centering
  \includegraphics[width=\columnwidth]{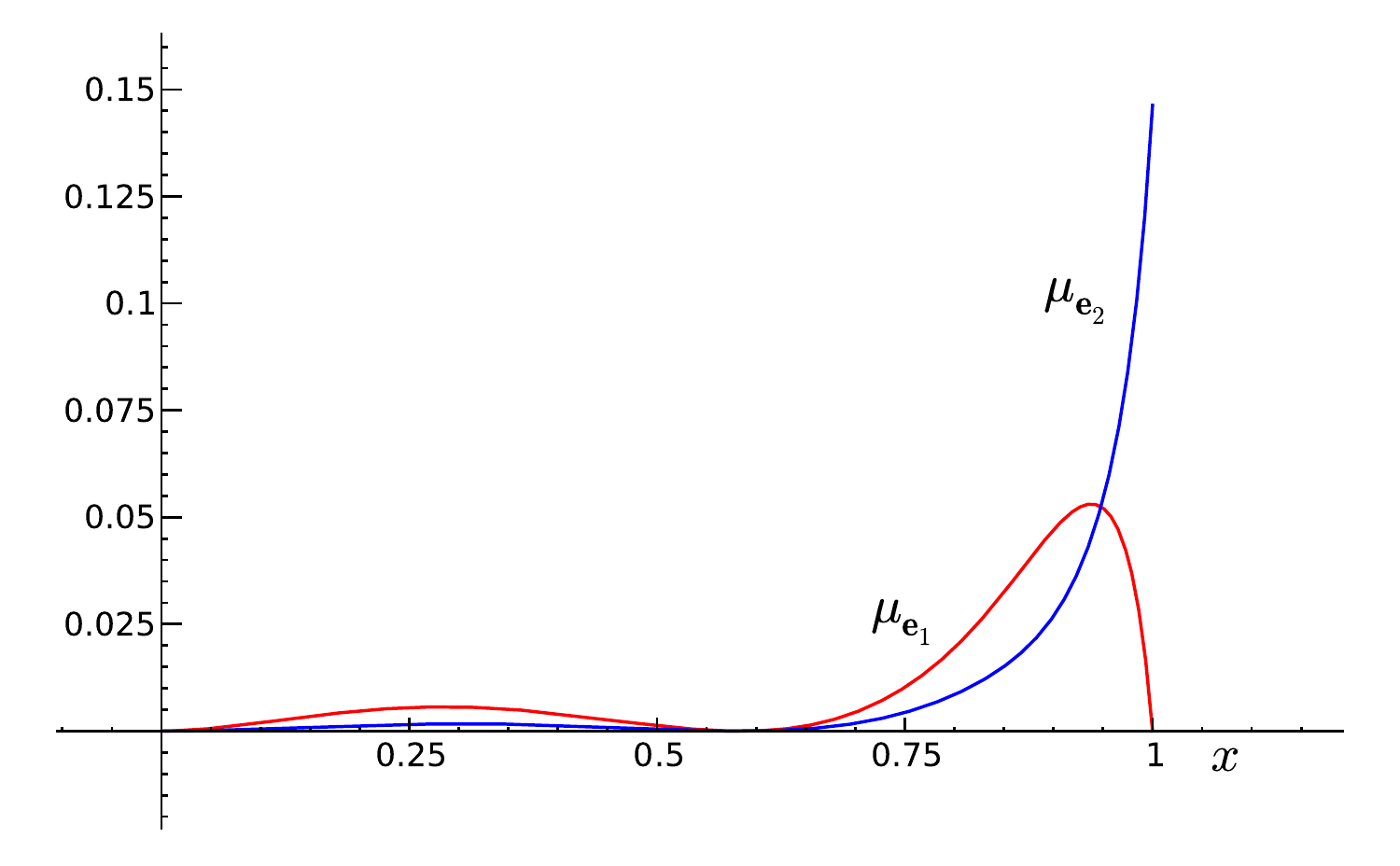}\\%
  \includegraphics[width=\columnwidth]{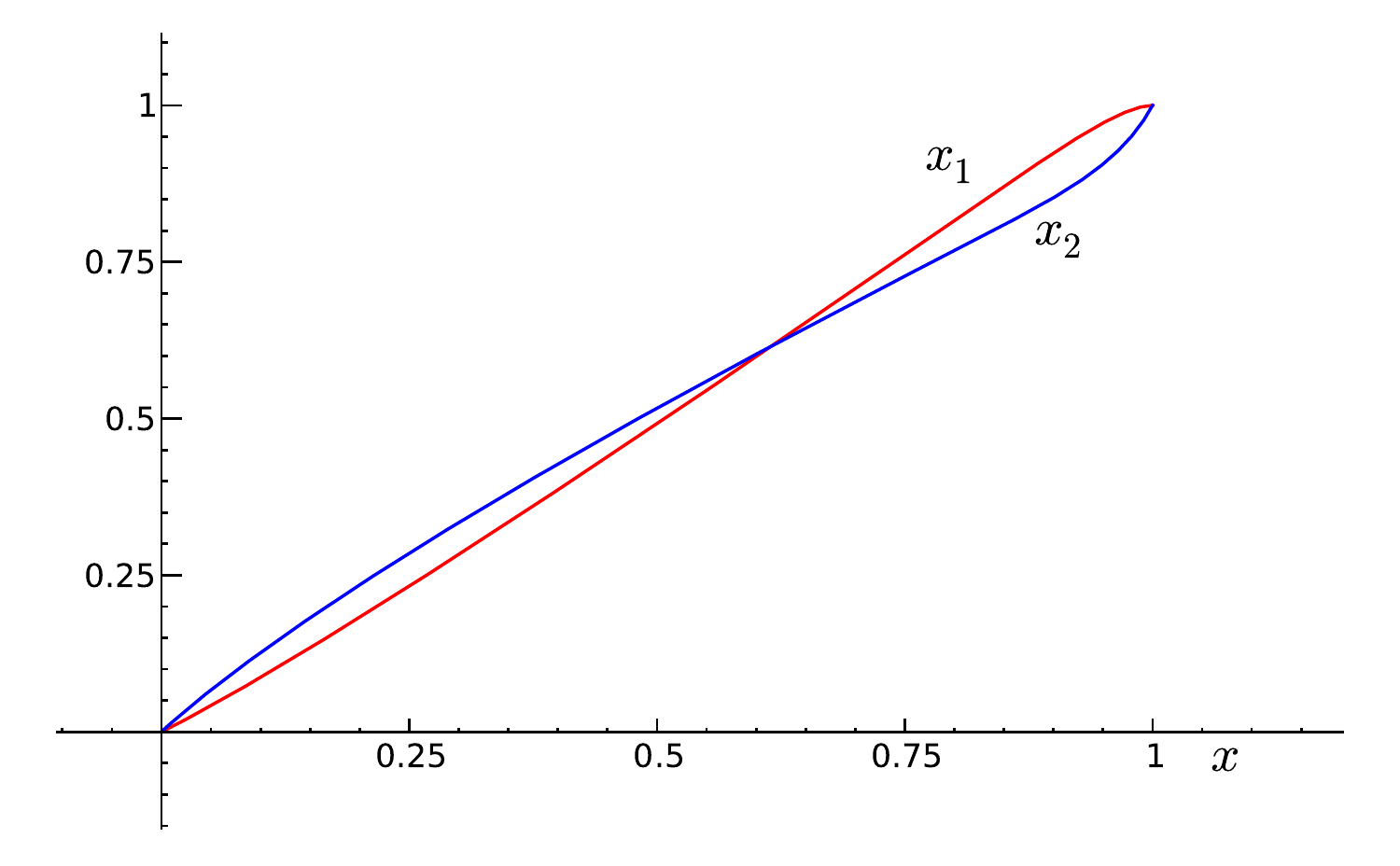}
  \caption{Example of peeling decoder state evolution for repeat-accumulate
    ensemble, systematic bits punctured, rate $1/3$.
    From~\cite{Richardson2002}, this ensemble has two edge types, $\nu(\vr,\vx)
    = r_1x_1^2 + \frac{1}{3}r_0x_2^3$ and $\mu(\vx) = x_1^2x_2$.  These plots
    follow the natural path for $\epsilon = 0.6175 \approx \epsilon^*$.  The
    horizontal axis for both plots is $x=\sum_i x_iE_i/E$, the weighted average
    of the elements of $\vx$.}
  \label{fig:met-pd-example}
\end{figure}

Below threshold (high probability of success), the differential equations for
$\vx(t)$ can be integrated numerically as shown in~\figref{fig:met-pd-example}.
Above threshold, all the degree-1 check node fractions go to zero for some
$\vx(t)$, so the natural schedule probabilities $\gamma_i$ (and hence the
equations for $\vx$) are indeterminate ($0/0$).

\appendix
\label{app:met-cn-soln}
In this section we present a derivation of the solution to the differential
equations describing the mean decoding trajectory for the peeling decoder
operating on an MET ensemble.  In the sequel, we denote $\gamma_i$ as a
function of $\vx$ (time) and $\vepsilon$ in accordance with
\secref{sec:met-xpath} where the choice of $\Gamma$ depends on time (through
$\vx$) and $\vepsilon$.  Consider the fraction of check nodes of degree greater
than one.  Denoting
\begin{align*}
  \lambda_j^\prime(\vepsilon,\vx) &= \ddt \lambda_j(\vepsilon,\vx)  
    = \sum_i \doverd[x_i]{t} \partialwrt{x_i} \lambda_j(\vepsilon,\vx),
\end{align*}
the expected number of ``other'' edges deleted simplifies to
\begin{align}
  \label{eqn:other-edges-soln}
  \tilde{a}_j(t) &= -\frac{E_j}{E} x_j \lambda_j^\prime(\vepsilon,\vx).
\end{align}
From the variable node solutions, the edge type fractions are
\begin{align}
  \label{eqn:ei-soln}
  e_i(\vepsilon,\vx) &= \frac{E_i}{E} x_i \lambda_i(\vepsilon,\vx).
\end{align}
Substituting these expressions into 
\eqref{eqn:dmu1}, the check node fractions follow
\begin{multline}
  \label{eqn:dmu-x}
  \ddt \mu_{\vd}(\vepsilon,\vx) \\
    = \sum_j \bigl[-(d_j+1)\mu_{\vd+\ve_j}(\vepsilon,\vx) + d_j\mu_{\vd}(\vepsilon,\vx)\bigr]
      \frac{\lambda_j^\prime(\vepsilon,\vx)}{\lambda_j(\vepsilon,\vx)}.
\end{multline}
Similar to~\cite{Luby1997}, the ``explicit'' solutions to
\eqref{eqn:dmu-x} are
\begin{multline}
  \label{eqn:mu-explicit}
  \hspace*{-0.5em}%
  \mu_{\vd}(\vepsilon,\vx) = \vlambda(\vepsilon,\vx)^{\vd} \biggl[c_{\vd} \biggr. \\ \biggl.
        - \sum_j (d_j+1) \int_{\tau=0}^t \mu_{\vd+\ve_j}(\vepsilon,\vx) 
        \frac{1}{\vlambda(\vepsilon,\vx)^{\vd}} 
        \frac{\lambda_j^\prime(\vepsilon,\vx)}{\lambda_j(\vepsilon,\vx)} 
        \,\intd\tau \biggr]\hspace{-0.4em}
\end{multline}
(inside the integral $\vx$ is a function of $\tau$ in place of $t$)
which admit the solutions
\begin{align}
  \label{eqn:mu-x}
  \mu_{\vd}(\vepsilon,\vx) 
    &= \sum_{\tilde{\vd} \geq \vd} (-1)^{\weight{\tilde{\vd}-\vd}} 
        \binom{\tilde{\vd}}{\vd} c_{\tilde{\vd}} \vlambda(\vepsilon,\vx)^{\tilde{\vd}}.
\end{align}
where $\weight{\vd}=\sum_i d_i$ is the $L_1$ norm.  Note that this sum ranges
over all larger check node degrees with non-zero time-dependent fractions.  The
constants $\{c_{\vd}\}$ are determined recursively from the initial conditions
as
\begin{align}
  \label{eqn:mu-soln}
  \mu_{\vd}(\vepsilon,\vx) 
    &= \sum_{\tilde{\vd} \geq \vd} \mu_{\tilde{\vd}} \binom{\tilde{\vd}}{\vd}
      \vlambda(\vepsilon,\vx)^{\vd} 
      \bigl[\vone - \vlambda(\vepsilon,\vx)\bigr]^{\tilde{\vd}-\vd}
\end{align}
for check nodes of degree greater than one.  
This solution obviously satisfies the initial conditions
\eqref{eqn:mu-initial}, and it can be shown that it also satisfies
\eqref{eqn:dmu-x}.

The fraction of degree-1 check nodes is the difference between the total
fraction of type-$i$ edges and the fraction of type-$i$ edges attached to
higher-degree check nodes.
\begin{align}
  \label{eqn:mu1-try1}
  &\hspace*{-2em}%
  \mu_{\ve_i}(\vepsilon,\vx) 
    = \dvavg e_i(\vepsilon,\vx) - \sum_{\vd \neq \ve_i} d_i\mu_{\vd}(\vepsilon,\vx)  
      \notag\\
    &= \dvavg e_i(\vepsilon,\vx) \notag\\
      &\quad- \sum_{\vd \neq \ve_i} d_i \sum_{\tilde{\vd} \geq \vd} \mu_{\tilde{\vd}} 
      \binom{\tilde{\vd}}{\vd} \vlambda(\vepsilon,\vx)^{\vd} 
      \bigl[\vone - \vlambda(\vepsilon,\vx)\bigr]^{\tilde{\vd}-\vd}  \hspace*{-2em}\notag\\
    &= \dvavg e_i(\vepsilon,\vx) \notag\\
      &\quad- \sum_{\tilde{\vd}} \mu_{\tilde{\vd}} \sum_{\vd \leq \tilde{\vd}} d_i
        \binom{\tilde{\vd}}{\vd} \vlambda(\vepsilon,\vx)^{\vd} 
        \bigl[\vone - \vlambda(\vepsilon,\vx)\bigr]^{\tilde{\vd}-\vd} \hspace*{-2em}\notag\\
      &\quad+ \sum_{\tilde{\vd}\geq\ve_i} \mu_{\tilde{\vd}} 
        \binom{\tilde{\vd}}{\ve_i} \vlambda(\vepsilon,\vx)^{\ve_i} 
        \bigl[\vone - \vlambda(\vepsilon,\vx)\bigr]^{\tilde{\vd}-\ve_i}
\end{align}
From definitions, the third term simplifies to $\dvidxavg{i}
\lambda_i(\vepsilon,\vx) \rho_i\bigl(\vone - \vlambda(\vepsilon,\vx)\bigr)$.
Furthermore, the second term is equal to $\sum_{\tilde{\vd}} \mu_{\tilde{\vd}}
\tilde{d}_i \lambda_i(\vepsilon,\vx) = \dvidxavg{i} \lambda_i(\vepsilon,\vx) =
\nu_i(\vepsilon,\vx)$.\footnote{Compare the generating functions $(\va +
  \vb)^{\tilde{\vd}}$ and $a_i \partialwrt{a_i} (\va + \vb)^{\tilde{\vd}}$ and
  their expansions with $\va=\vlambda(\vepsilon,\vx)=\vone-\vb$.}  Combining
these simplifications with $\dvavg e_i(\epsilon,\vx) = \dvidxavg{i} x_i
\lambda_i(\vepsilon,\vx)$, we have the solution~\eqref{eqn:mu1-soln} given in
Lemma~\ref{lem:mu1-soln}.


\bibliographystyle{IEEEtran}
\bibliography{IEEEabrv,../../rwhabrv,../../rwhinton}

\end{document}